\documentstyle[12pt]{article}
\begin{document}
\font\ninerm = cmr9

\def\footnoterule{\kern-3pt \hrule width \hsize \kern2.5pt}

\pagestyle{empty}

\begin{flushright}
hep-ph/0006210 \\
$~$ \\
June 2000
\end{flushright}

\vskip 0.5 cm

\begin{center}
{\large\bf Cosmic rays and TeV photons 
as probes of \\
quantum properties of space-time}
\end{center}
\vskip 1.5 cm
\begin{center}
{\bf Giovanni AMELINO-CAMELIA}$^a$ and {\bf Tsvi PIRAN}$^b$\\
\end{center}
\begin{center}
{\it $^a$Dipart.~Fisica,
Univ.~Roma ``La Sapienza'',
P.le Moro 2, 00185 Roma, Italy}\\
{\it $^b$Racah Institute of Physics, Hebrew University,
Jerusalem 91904, Israel}
\end{center}

\vspace{1cm}
\begin{center}
{\bf ABSTRACT}
\end{center}

{\leftskip=0.6in \rightskip=0.6in 
It has been recently observed that small violations of
Lorentz invariance, of a type which may arise in quantum gravity,
could explain both the observations
of cosmic rays above the GZK cutoff and the observations of 20-TeV
gamma rays from Markarian 501.  We show here that different pictures
of the short-distance structure of space-time would lead to different
manifestations of Lorentz-invariance violation. Specifically,
the deformation of Lorentz invariance needed to resolve these
observational paradoxes can only arise within commutative
short-distance pictures of space-time. In noncommutative space-times
there is no anomalous effect, at least at leading order.     
Also exploiting the fact that arrival-time delays between
high energy photons with different energies would arise in both
the commutative and the noncommutative Lorentz-violation pictures,
we describe an experimental programme, based on time-of-arrival
analysis of high energy photons and searches of violations of GZK
and TeV-photon limits, which could discriminate between alternative
scenarios of Lorentz-invariance breakdown and could provide and
unexpected window on the (quantum) nature of space-time
at very short distances.       
}

\newpage
\baselineskip 12pt plus .5pt minus .5pt
\pagenumbering{arabic}
\pagestyle{plain} 


A rather robust expectation that is emerging from
theory work on short-distance (so called, ``quantum gravity")
properties of space-time is that space-time symmetries
may require modification.
In particular, quantum-gravity effects inducing
some level of nonlocality or noncommutativity
would affect even the most basic flat-space
continuous symmetries,
such as Lorentz invariance.
This has been recently emphasized in various quantum-gravity
approaches~[1-14]
based on critical or
noncritical string theories,
noncommutative geometry, canonical quantum gravity.
While we must be open to the possibility that
some symmetries be completely lost, it appears plausible
that some of them be
not really lost but rather replaced by
a Planck-scale-deformed version,
and some mathematical frameworks which could consistently
describe such deformations have emerged in the mathematical-physics
literature~\cite{kpoinap,majrue,gackpoinplb,lukipap,gacmaj}.

We shall here focus on quantum-gravity motivated
violations of Lorentz invariance, but of course
Lorentz-invariance violations do not have to be
necessarily associated with quantum gravity
and in fact even outside the quantum-gravity literature
there is a large amount of work on the theory and
phenomenology of violations of Lorentz invariance
(see, {\it e.g.}, the recent Refs.~\cite{colgla,jackost},
which also provide a good starting point for a literature search
backward in time).

Since quantum-gravity formalisms are still too complex to allow
definite quantitative predictions of the effects,
possible deformations of Lorentz invariance have been
investigated through phenomenological models. In particular,
the investigation of the Lorentz-violating
class of dispersion relations 
\begin{equation}
E^2 - \vec{p}^2 \simeq 
\xi_{\pm} E^2 \left({E \over E_{QG}}\right)^{n}
\label{dispone}
\end{equation}
was proposed, for the case $n=1$,
in Refs.~\cite{aemn1,gackpoinplb,grbgac} and,
for all $n$, in Ref.~\cite{polonpap}.
In (\ref{dispone}) $E$ and $\vec{p}$ 
denote the energy and the (3-component) momentum of the particle,
$\xi_{\pm}$ sets the overall sign
of the deformation ({\it i.e.} $\xi_{\pm} = \pm 1$),
$E_{QG}$ is a fixed quantum-gravity energy scale
(which is of course not yet known but can be suspected to be
in the neighborhood of the Planck scale $10^{19} GeV$)
and $n$ is a fixed quantum-gravity number
(which in principle could take any value, but
we will assume it to be an integer,
as it appears to be plausible~\cite{polonpap})
that specifies how strongly the magnitude of the deformation
is suppressed by $E_{QG}$.
In the following we will refer to these scenarios using the
notation $n_{sign(\xi_{\pm})}$, leaving implicit the dependence
on $E_{QG}$ (for example, we refer to the
case with $\xi_{\pm} = -1$ and $n=1$ as the case $1_{-}$,
and within the case $1_{-}$ we consider the range of
acceptable values of $E_{QG}$).

Early phenomenological interest in the proposal
(\ref{dispone}) came from studies
based on time-of-arrival
analyses~\cite{grbgac,schaef,billetal}
of photons associated with gamma-ray bursts
or with Markarian 501.

We are here primarily concerned with another class of implications
of violations of Lorentz invariance which has been more recently
emphasized in the literature.
It has been
observed~[17,21-26]
that evidence for Lorentz-invariance violation may arise
from analyses of the interactions of cosmic rays with the
cosmic microwave background radiation and from analyses of
the interactions
of high-energy gamma-rays with the universal infrared background.
In both cases we observe on earth particles with energies above a
critical treshold for interaction with background radiation
(cosmic microwave background radiation 
for ultra-high-energy cosmic rays
and infrared background for 10-TeV photons).
Lorentz-invariance violation could change these thresholds.
Concerning the proposal (\ref{dispone}) relevant analyses were
given in Refs.~\cite{kifu,kluz,ita,sato,aus}.
The basic point of these studies is that the anomalous effects
encoded in the $E_{QG}^{-n}$ term of (\ref{dispone}), while
being safely negligible in other contexts, can become very
significant when a low-energy background photon collides with a
high-energy particle with momentum close to the one required by
threshold pion production or electron-positron pair creation.

We summarize this argument for the case of photon-photon collisions
(the analysis of proton-photon pion production relevant
for the cosmic-ray context is completely analogous, although
of course less symmetrical).
We start by discussing Lorentz-invariant
head-on collision between
a soft photon of energy $\epsilon$ and momentum $q$
and a high-energy photon of energy $E$ and momentum $p$.
This collision may create an electron and a positron
both basically moving along the original direction of travel
of the hard photon.
Denoting with $E_+$ ($E_-$) and $p_+$ ($p_-$)
the energy and momentum of the emerging positron (electron),
energy conservation
and momentum conservation imply
\begin{equation}
E+\epsilon=E_++E_-
\label{econsv}
\end{equation}
\begin{equation}
p-q=p_++p_-
\label{pconsv}
\end{equation}
Denoting with $p'$ the common modulus of $p_+$ and $p_-$,
and using the conventional
Lorentz-invariant relation between energy and momentum
one then obtains the relations
\begin{equation}
E+\epsilon=p+\epsilon~,~~~p-q=p-\epsilon~,~~~
E_+ + E_-= 2 \sqrt{p'^2+m^2}
\simeq 2 p' + {m^2 \over p'}
~,~~~p_++p_-=2p' ~,
\label{lirel}
\end{equation}
where $m$ denotes the electron mass and the fact that $p$
(and, as a consequence, $p'$) is a large momentum
has been used to approximate the square root.
Combining (\ref{econsv}), (\ref{pconsv})
and (\ref{lirel}) one easily obtains that the $p$-threshold
for electron-positron pair creation is
\begin{equation}
p \ge p_{th} = {m^2 \over \epsilon} ~.
\label{lithresh}
\end{equation}
This of course is easily rederived by analyzing the photon-photon
system in the C.O.M. frame, where both photons carry momentum
$\sqrt{E \epsilon}$ and at threshold the 
electron-positron pair is produced at rest.

This standard Lorentz-invariant analysis may be affected
by the type of deformations codified
in (\ref{dispone}).
This has been analyzed in Refs.~\cite{kifu,kluz,ita,sato,aus}
within two main assumptions: (i) Eq.~(\ref{dispone})
holds in the ``laboratory frame''
(ii) when a noncommutative geometry was used to support
the scenario it was assumed that the
noncommutativity would only affect the analysis
through Eq.~(\ref{dispone}).
To illustrate in a specific example the basic ingredients
of the analyses reported in Refs.~\cite{kifu,kluz,ita,sato,aus},
let us consider the case $1_-$, in which
Eq.~(\ref{dispone}) predicts a deformed
relation between energies and momenta such that
within the assumptions (i),(ii)
Eq.~(\ref{lirel}) would be replaced by
\begin{equation}
E+\epsilon=p+\epsilon - {p^2 \over 2 E_{QG}}
~,~~~p-q=p-\epsilon~,~~~
E_+ + E_- \simeq 2 p' + {m^2 \over p'} - {p'^2 \over E_{QG}}
~,~~~p_++p_-=2p' ~,
\label{lv1rel}
\end{equation}
where we only included the leading corrections (terms
suppressed by both the smallness of $E_{QG}^{-1}$ and the
smallness of $\epsilon$ or $m$ were neglected).
Combining (\ref{econsv}), (\ref{pconsv})
and (\ref{lv1rel}) one obtains a deformed equation
describing the $p$-threshold
for electron-positron pair creation:
\begin{equation}
p_{th,1_{-}} = {m^2 \over \epsilon}
+ {p_{th,1_{-}}^3 \over 8 \epsilon E_{QG}} 
~.
\label{lithresh2}
\end{equation}
Analogous relations are obtained~\cite{ita} for the
scenarios $1_+$, $2_-$, $2_+$; for example, $2_-$ leads to 
\begin{equation}
p_{th,2_{-}} = {m^2 \over \epsilon}
+ {3 p_{th,2_{-}}^4 \over 16 \epsilon E_{QG}^2} 
~.
\label{lithresh2b}
\end{equation}

The fact that the scale $E_{QG}$ is believed to be very high might
give the erroneous impression that the new
term $p_{th,1_{-}}^3 / (8 \epsilon E_{QG})$
present in Eq.~(\ref{lithresh2}) (valid in the $1_-$
scenario, which we continue to use as our illustrative case) could
always be safely neglected, but this is not the
case~\cite{kifu,kluz,ita,sato,aus}.  For given high value of $E_{QG}$
one finds low enough values of $\epsilon$ for which the ``threshold
anomaly'' (displacement of the threshold) is significant. Actually in
this case $1_-$ (as well as in the case $2_-$) there are values of
$\epsilon$ that are low enough to cause the disappearance of the
corresponding threshold (values of $\epsilon$ for which
Eq.~(\ref{lithresh2}) has no solutions).  Indeed, assuming
(\ref{lithresh2}) one would predict dramatic departures from the
ordinary expectations of Lorentz invariance; in particular, if $E_{QG}
\sim 10^{19}GeV$, according to (\ref{lithresh2}) one would expect,
contrary to the predictions of a Lorentz-invariant treatment, that the
Universe be transparent to TeV photons~\cite{kluz}, while the
corresponding result obtainable in the cosmic-ray context would imply
that the GZK cutoff may be violated~\cite{kifu}.
(An analysis of violations of the GZK cutoff within other
schemes of Lorentz-invariance violation can be found
in Ref.~\cite{colgla}.)
Remarkably these predictions appear
to fit well the preliminary indications of some recent
puzzling cosmic-ray observations~\cite{kifu} and observations of
astrophysical high-energy photons~\cite{kluz,aus}.

Concerning the significance of the phenomenological
scenario $1_-$ we stress here that this is (to our knowledge)
the only proposal that would allow to describe simultaneously
both Universe transparency to TeV photons
and violations of the cosmic-ray GZK cutoff.
This point, which was not stressed by previous authors,
may render the proposal (\ref{dispone})
quite appealing if the preliminary
indications of transparency to TeV photons and 
violations of the GZK cutoff turn out to be confirmed by
more robust data.

Also relevant for the phenomenological analysis (but not
discussed in previous studies) is the fact that the small
quantum-gravity correction becomes significant only
when very close to the special conditions for the threshold.
Analyses of more ordinary collisions are left basically
unaffected by the deformation. This is important since in principle
Lorentz invariance plays a role both in the calculation of the
threshold and in the algorythms used to determine the measured
photon or cosmic ray energy, and therefore one might wonder whether
the Lorentz-violations scenarios here considered could affect
our determination of the energies as significantly as it affects
the analysis of the threshold. This is indeed not the case:
while we find that the analysis of the threshold can be
significantly modified by the deformation, it is easy to check that
instead the minute quantum-gravity deformation which we are
considering would not significantly affect the processes
used to determine the energies of the hard particles.

Besides contributing these elements of analysis relevant
for the development of the phenomenology here under consideration,
another objective
of the present Letter is the one of further exploring
the role that predictions of the type (\ref{lithresh2}),
(\ref{lithresh2b}),
which can follow from Eq.~(\ref{dispone}),
may have on the development of an experimental programme for
quantum gravity~\cite{ehns,grbgac,gacgwi,ahlunature,polonpap},
particularly as an opportunity to differentiate between
alternative short-distance pictures of space-time.
With respect to this last point it is important to reanalyze
the two mentioned assumptions (i),(ii)
on which the result (\ref{lithresh2})
relies. We use
the specific example of the simple ``$\kappa$-Minkowski''
noncommutative space-time
developed in Refs.~\cite{kpoinap,majrue,gackpoinplb,lukipap,gacmaj}
to illustrate the implications
of a consistent analysis within noncommutative geometry and/or
of a prescription for describing the phenomenon in frames
other than the laboratory frame.

One key point is that in $\kappa$-Minkowski
a relation of type (\ref{dispone})
can be obtained as a direct consequence of the
$\kappa$-Poincar\'e
invariance~\cite{kpoinap,majrue,gackpoinplb,lukipap,gacmaj}
of this space-time.
$\kappa$-Poincar\'e is a deformation of the Poincar\'e group
in which precise rules are still available for
the description of changes of frame of reference,
and actually in a $\kappa$-Minkowski space-time
Eq.~(\ref{dispone}) can characterize significantly these
rules as an invariant.
Another key point is that consistency with the
noncommutative nature of $\kappa$-Minkowski space-time
requires~\cite{kpoinap,majrue,gackpoinplb,lukipap,gacmaj}
that the law of addition of momenta be accordingly modified.
The physical interpretation of these deformed laws of
addition is still being developed, but there is now
a definite prescription put forward in Ref.~\cite{gacmaj}.
Importantly, in the (laboratory-frame) context here of interest
the prescription of Ref.~\cite{gacmaj} leaves basically
unaffected (in leading order) the sum of the two very different
momenta $p$ and $q$ while the sum
of the two momenta $p_+$ and $p_-$, which are
roughly parallel and roughly
of the same magnitude, is modified in leading order:
\begin{equation}
p - q \simeq p-q~,~~~p_++p_-=2p' + {p'^2 \over E_{QG}}  ~.
\label{psum}
\end{equation}
The fact that $p'=p_+ = p_-$
but $p_+ + p_- = 2 p' + p'^2 / E_{QG} \neq 2 p'$
reflects the fact that~\cite{gacmaj} the noncommutativity
of $\kappa$-Minkowski space-time requires its associated
energy-momentum space to have nontrivial
(curved, nonabelian) geometry.

In light of these considerations it is easy to see that
in $\kappa$-Minkowski it is the ordinary
threshold equation
(\ref{lithresh}), rather than the deformed
threshold equation
(\ref{lithresh2}), that follows from the
case $n=1$, $\xi_{\pm} = -1$ in Eq.~(\ref{dispone}).
This can be seen both in the laboratory
frame and in the C.O.M. frame.
In the laboratory frame the result follows from the fact
that the deformed
law of addition of momenta requires one to replace
(\ref{lv1rel}) with
\begin{equation}
E+\epsilon=p+\epsilon - {p^2 \over 2 E_{QG}}
~,~~~p-q=p-\epsilon~,~~~
E_+ + E_- \simeq 2 p' + {m^2 \over p'} - {p'^2 \over E_{QG}}
~,~~~p_++p_- \simeq 2p' + {p'^2 \over E_{QG}} 
\label{lv2rel}
\end{equation}
and this combines with (\ref{econsv}) and (\ref{pconsv})
in a way that gives rise to two equal-magnitude but opposite
corrections to the threshold equation, thereby giving
us back the original threshold equation (\ref{lithresh})
(the one that also follows from assuming ordinary Lorentz
invariance).
In the C.O.M.~frame the same cancellation appears
under a different guise: In the C.O.M.~frame the two photons
have the same energy $E_{com}$, which in the $\kappa$-Minkowski 
framework, where (\ref{dispone})
with $n=1$, $\xi_{\pm} = -1$ would be a genuine invariant,
must satisfy the leading-order relation
\begin{equation}
(E+\epsilon)^2 - (p-q)^2 
+ {E^3 \over E_{QG}}
\simeq 4 E_{com}^2  ~,
\label{disptwo}
\end{equation}
where we also used the fact that,
according to the prescription of Ref.~\cite{gacmaj},
there is no leading-order
deformation of the sum of two momenta of the same
magnitude and opposite direction.
Using $q \simeq \epsilon$ and $E \simeq p - p^2 /(2 E_{QG})$,
Eq.~(\ref{disptwo}) leads to the result
\begin{equation}
E_{com} \simeq \sqrt{p \epsilon}  ~,
\label{ecom}
\end{equation}
just as in the ordinary Lorentz-invariant case.

We therefore conclude that $\kappa$-Minkowski
noncommutative space-time provides us an example
of space-time in which Eq.~(\ref{dispone}) holds
but the new threshold-related
effects generically attributed
to Eq.~(\ref{dispone}) in Refs.~\cite{kifu,kluz,ita,sato,aus}
are not present.
This may have wider validity: it appears likely that
in any noncommutative space-time in which
Eq.~(\ref{dispone}) holds and corresponds to an invariant
of the theory the mentioned threshold-related
effects would be absent.
On the contrary it appears that in a commutative space-time
(in which one would be surprised to
encounter $\kappa$-Poincar\'e-type deformations of the law of
addition of momenta)
the validity of Eq.~(\ref{dispone}) would inevitably lead
to the threshold-related
effects discussed here and
in Refs.~\cite{kifu,kluz,ita,sato,aus}.
Noticeably, in such space-times Eq.~(\ref{dispone})
could only be valid in one preferred class of reference frames
(which some authors~\cite{colgla} tentatively identify
with the frame in which the cosmic microwave background
is isotropic), and a major challenge appears to be the development
of a description of the new phenomena in the C.O.M. frame (which
one might still expect to exist, although perhaps connected
in highly nontrivial manner to the laboratory frame).

The fact that deformations of type (\ref{dispone})
should cause the threshold-related effects
discussed here and in Refs.~\cite{kifu,kluz,ita,sato,aus}
if space-time is commutative and should not cause them
if space-time is noncommutative 
opens up an opportunity to distinguish experimentally
between different short-distance pictures of space-time.
In fact, the information obtainable with these
threshold-related investigations nicely complements
the fact that in any space-time (commutative or not)
that supports the dispersion relation (\ref{dispone})
with $n=1$
one expects the time-delay effects discussed in
Refs.~\cite{grbgac,schaef,billetal}.
If this time delays are not observed in
near-future upgrades of the experimental
programme of observation of gamma-ray
bursts~\cite{grbgac,grbrev}
the scenario $1_-$ will be completely ruled out.
If instead the time delayes are observed
we will have evidence that the dispersion relation
(\ref{dispone}) is verified in Nature, and we will then be able
to establish whether the space-time structures
responsible for (\ref{dispone}) are commutative or noncommutative
by analyzing the threshold-related effects
discussed here and in Refs.~\cite{kifu,kluz,ita,sato,aus}
(also on these effects experimental limits should rapidly
improve in the coming years~\cite{ita,aus}).

In the wider picture of quantum-gravity research these
are very significant developments, at least in as much as
they show that gaining experimental insight on
Planck-scale-related physics is no longer impossible.
The fact that at least some level of
Planck-scale sensitivity could be achieved in some contexts
had already emerged from the analyses reported
in Refs.~\cite{ehns,grbgac,gacgwi,ahlunature,polonpap}.
We are finding that
in the context of the Planck-scale-deformed
dispersion relation (\ref{dispone}) and the types of
space-times that could support it
the experimental studies that are becoming possible by combining
the analyses reported in Refs.~\cite{grbgac,schaef,billetal}
and the analyses reported here and
in Refs.~\cite{kifu,kluz,ita,sato,aus}
actually constitute a rather robust experimental programme
of investigation.

We close observing that (especially in light of the fact that
the ambitious ``quantum-gravity problem''  may still
require several decades of theoretical and experimental study)
it may eventually become significant that
the type of threshold-related effects
discussed here and in Refs.~\cite{kifu,kluz,ita,sato,aus}
could in principle allow dedicated controlled experiments.
For example, even the deformed threshold equation
(\ref{lithresh2b}), which can result from the scenario $2_-$
and is quadratically suppressed by $E_{QG}$,
predicts a shift of the threshold of a few percent
for soft photons with, say, $\epsilon \sim 3 \cdot 10^{-6}eV$.
This implies that studies of the cross section
for collisions between soft
photons with $\epsilon \sim 3 \cdot 10^{-6}eV$
and hard photons with energies in the neighborhood
of $E \sim m^2/\epsilon \sim 10^8 GeV$
could test (\ref{lithresh2b}).
We are probably very far (several decades?)
from being able to devise
this type of experiments, but clearly these are
much closer to
the realm of experiments doable in the foreseeable
future than the traditionally-considered class of
gedanken quantum-gravity experiments,
the only one mentioned in most quantum-gravity reviews
until a few years ago, which relied on the study
of collisions of particles endowed with Planckian energies,
{\it i.e.} with energies in the neighborhood of $10^{19}GeV$.

\bigskip
One of us (G.A.-C.) greatfully acknowledges stimulating
conversations with Daniele Fargion and Fedele Lizzi.

\baselineskip 12pt plus .5pt minus .5pt

\end{document}